\documentclass[pre,aps,twocolumn,superscriptaddress,showpacs,floatfix]{revtex4}
\usepackage{graphicx}
\usepackage{subfigure}
\usepackage{amssymb}

\begin{document}

\title{Pore formation in fluctuating membranes}

\author{Oded Farago}
\email{farago@mrl.ucsb.edu}
\affiliation{Materials Research Laboratory, 
University of California, Santa Barbara, 93106}
\altaffiliation{Also at Department of Physics, Korea Advanced Institute of
Science and Technology (KAIST), 373-1 Kusong-dong, Yusong-gu, Taejon
305-701, South Korea.}
\author{Christian D. Santangelo}
\email{santa@mrl.ucsb.edu}
\affiliation{Department of Physics, 
University of California, Santa Barbara, 93106}
\date{\today}

\begin{abstract}
  We study the nucleation of a single pore in a fluctuating lipid
  membrane, specifically taking into account the membrane
  fluctuations, as well as the shape fluctuations of the pore.  For
  large enough pores, the nucleation free energy is well-described by
  shifts in the effective membrane surface tension and the pore line
  tension.  Using our framework, we derive the stability criteria for
  the various pore formation regimes. In addition to the well-known
  large-tension regime from the classical nucleation theory of pores,
  we also find a low-tension regime in which the effective line and
  surface tensions can change sign from their bare values. The latter
  scenario takes place at sufficiently high temperatures, where the
  opening of a stable pore of finite size is entropically favorable.
\end{abstract}


\maketitle

\section{Introduction}

Lipid bilayers play an important role in living cells as barriers
separating the inside of the cell from the extracellular environment,
as well as segregating the cell into separate internal
compartments~\cite{alberts}.  A vital feature of those membranes is
the ability to remain intact under variety of external
perturbations~\cite{dimitrov}.  At the same time, however, many
cellular processes, including endo- and exocytosis, lysis, and cell
signaling require breaking the membrane structure and forming a
non-bilayer transient, such as a membrane
pore~\cite{chernomordik,marsh,schmidt,bechinger}. The opening of
stable pores in biological membranes is also an important step for
drug delivery~\cite{drugdelivery} and gene therapy~\cite{genetherapy}.
Consequently, much attention has been focused on understanding the
processes leading to the formation of pores and the mechanisms
controlling their stability.

Two types of pores dominate membrane permeability: free lipid pores
and peptide-lined pores~\cite{lipowsky_sackmann}. The interest in
lipid pores has greatly increased in the past few years with the
development of new experimental techniques to induce and study pore
formation in biomimetic, single-component, lipid membranes. One
approach to nucleate a pore is known as electroporation, where
an electric field that produces compressive stress is used to disrupt the
membrane~\cite{abidor,needham,wilhelm,winterhalter,bermudez}. Holes in
lipid membranes have also been opened by other methods, including
intense illumination~\cite{barziv,sandre}, suction through a
micropipette~\cite{olbrich}, adhesion on porous or decorated
substrates~\cite{guedeau,singhvi}, and osmotic
swelling~\cite{ertel,mui}.

Most theories of pore formation to date derive from a model based on
classical nucleation theory~\cite{litster}. The model conceives the
membrane as a two dimensional elastic medium characterized by a free
energy per unit area (``surface tension'') $\sigma$. The formation of
a circular hole of radius $r_0$ is driven by the reduction in the
tension energy $-\sigma\pi r_0^2$ and is opposed by an edge energy
proportional the pore perimeter $\Gamma 2\pi r_0$, where $\Gamma$, the
line tension, denotes the energy per unit length along the pore's rim.
The net energy is, thus, given by
\begin{equation}
E=\Gamma\,2\pi r_0-\sigma\pi r_0^2.
\label{nucleation}
\end{equation}
Assuming $\Gamma>0$ and $\sigma>0$, Eq.(\ref{nucleation}) predicts
that a pore with a radius larger than the critical value of
\begin{equation}
r_0>\frac{\Gamma}{\sigma}
\label{criticalradius}
\end{equation}
is unstable in the sense that it will grow without bound and,
ultimately, will rupture the membrane. Such a large pore will be
created only if the nucleation energy barrier 
\begin{equation}
\Delta E=\frac{\pi\Gamma^2}{\sigma}
\label{barrier}
\end{equation}
is accessible by thermal fluctuations. For typical estimates of the
line tension, $\Gamma\sim 10^{-6}\ {\rm
  dyn}$~\cite{zhelev,moroz,karatekin}, thermally driven rupture
requires a surface tension on the order of $1\ {\rm dyn}/{\rm cm}$.

The problem with the above model is that it precludes the existence of
stable pores of finite size. Nevertheless, long-lived pores that remained
open for several seconds before resealing have been observed in
experiments~\cite{bermudez,sandre,zhelev,moroz,karatekin}. Opening of
transient pores has been also reported is several computer
simulations~\cite{muller,groot,farago,stecki}. This has led people to
reexamine the basic assumptions underlying Eq.(\ref{nucleation}). Improved
theoretical models succeeded in explaining the formation of stable or
long-lived metastable pores by considering the fact that, once pores have
been nucleated, their further opening is expected to relax the surface
tension~\cite{moroz,farago,netz,brochardwyart,talanquer,fournier,levin}. In
the case of a planar membrane, it is the increase of the area density of the
lipids (occurring concurrently to the dilation of the pore) that reduces the
mechanical tension. For vesicles, the opening of a pore allows the internal
contents to escape, which reduces the osmotic pressure and the associated
Laplace tension.

Most of the theoretical models have so far, however, neglected the effect of 
membrane fluctuations on the opening and thermodynamic stability of pores.
%, in the sense that the
%membranes (vesicles) have been usually assumed to be flat (spherical)
The entropy associated with the shape of the pore has mostly been
ignored also.
%, and only circular pores that (for a given pore area) minimize
%the line tension energy have been allowed. 
%Some exceptions to this rule exist, however. 
A few recent studies of the entropic contribution to the free energy
of nucleating a pore, have led to some new interesting predictions.
The most remarkable result has been obtained by Shillcock and
Boal~\cite{shillcock_boal} in computer simulations of two-dimensional
fluid tethered surfaces.  They found that pores appeared at zero, and
even small negative surface tension. Their interpretation of this
surprising finding was that entropy, which favors the formation of
non-circular pores, reduces the effective line tension of the pore and
makes it negative at sufficiently high temperature. An entirely
different fluctuation effect has been discussed by Sens and
Safran~\cite{sens_safran} who considered circular pores, but allowed
membrane fluctuations. Their study suggests that positive stress must
be applied in order to facilitate the opening of a pore in a
fluctuating membrane, and that the nucleation barrier for pore
formation is too high to be overcome by thermal fluctuations. More
recently, we~\cite{bhgdead1} have demonstrated that the primary effect
of membrane fluctuations on circular holes is to reduce the effective surface tension,
thereby making the opening of a pore more difficult in comparison to
the zero-temperature case.

In this paper, we carry out a statistical mechanical analysis of pore
formation in bilayer membranes. Membrane elasticity is described by
the Helfrich Hamiltonian, which includes the curvature energy and a
surface tension term.  A line tension term is introduced to account
for the energetic penalty at the pore edge. We calculate the free
energy for nucleating a single pore, systematically taking into
account both membrane fluctuations and the entropy due to pore shape.
We show that, for large enough pores, the pore nucleation free energy
takes the form of Eq.(\ref{nucleation}), with the surface and line
tensions replaced by their effective (renormalized) values,
$\sigma_{\rm eff}$ and $\Gamma_{\rm eff}$. The effective coefficients
are usually smaller than the bare counterparts and, at high enough
temperatures, may even be negative. When $\Gamma_{\rm eff}<0$, the
opening of a stable pore becomes entropically favored and may occur in
weakly stressed membranes. The size of a thermal pore can be varied by
changing the tension applied on the membrane, and the membrane is
ruptured when $\sigma_{\rm eff}>0$.

The paper is organized as follows: The system Hamiltonian is constructed
in section \ref{section_derivation}. We show that Eq.(\ref{nucleation})  
for the energy should augmented by terms representing corrections due to
the pore's shape and membrane's height fluctuations. Tracing over the
relevant variables yields the corresponding free energy. The derivation of
the free energy is presented in sections \ref{section_surface} and
\ref{section_line}, in which the thermal corrections to the surface and
line tensions are calculated. In section \ref{section_summary} we discuss
our results and suggest some possible further extensions of the present
study.

\section{Derivation of the Hamiltonian}
\label{section_derivation}

We consider a bilayer membrane consisting of $N$ lipids that spans a
planar circular frame of total area $A_p=\pi L_p^2$, in which a 
quasi-circular pore has been formed. For a nearly flat membrane with
arbitrary parametrization $\mathbf{X}(x1,x2)$, the Helfrich
Hamiltonian is given by
\begin{equation}
\label{eq:HelfrichEnergy}
{\cal H} = \int_M dx_1\!dx_2\,\sqrt{g}\left[\sigma+ \frac{\kappa}{2} H^2 
+ \bar{\kappa}K\right],
\end{equation}
where $g$ is the determinant of the metric tensor $g_{\alpha \beta} =
\partial_\alpha \textbf{X} \cdot \partial_\beta \textbf{X}$, while $H$
and $K$ denote the total and Gaussian curvatures, respectively. The
elastic coefficients appearing in the Helfrich Hamiltonian are the
surface tension $\sigma$, the bending rigidity $\kappa$, and the
Gaussian rigidity $\bar{\kappa}$. We assume that the bilayer membrane
is symmetric with no spontaneous curvature.  The integration in
Eq.(\ref{eq:HelfrichEnergy}) is carried over the two-dimensional
manifold $M$, representing the surface of the membrane.

\subsection{Gaussian curvature}

\begin{figure}[h,t]
\begin{center}
\scalebox{1.05}{\centering \includegraphics{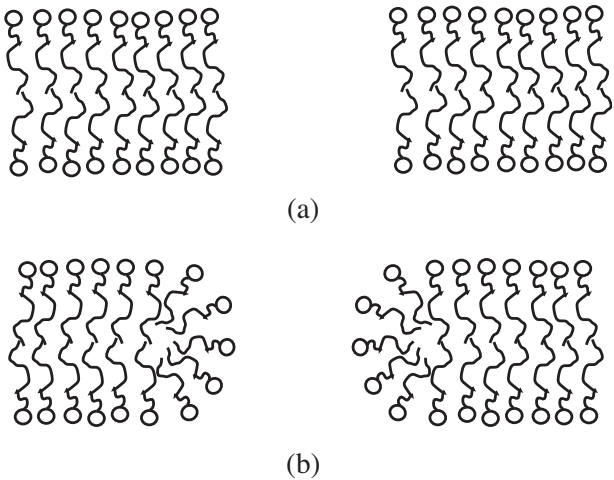}}
\end{center}
\caption{\label{fig:inclusion_pore} Schematic microscopic models for 
  hydrophobic (a) and hydrophilic (b) pores. For a hydrophilic pore,
  the boundary actually joins the two monolayer of the lipid
  membrane.}
\end{figure}

Understanding the contribution of the Gaussian curvature term [last
term in Eq.(\ref{eq:HelfrichEnergy})] to the free energy requires
looking at the structure of the membrane on the molecular level. Two
distinct models of pores have been discussed in the literature and are
shown schematically in fig.~\ref{fig:inclusion_pore}.
Fig.~\ref{fig:inclusion_pore} (a) depicts a cylindrical pore where the
lipids in the vicinity of the pore remain oriented parallel to the
membrane surface. Such a pore is called ``hydrophobic'', and the
origin of the pore line tension is the energy due to the exposure of
the tails of the lipids at pore's rim to water. The other case is of a
``hydrophilic'' pore, shown in fig.~\ref{fig:inclusion_pore} (b),
where the lipids curve at the pore's rim thus shielding their
hydrophobic parts from the aqueous contact. The line tension of a
hydrophilic pore is due to the curvature energy involved in the
reorientation of end molecules.

The Gaussian curvature term in the Helfrich Hamiltonian is calculated
differently for hydrophobic and hydrophilic pores. For hydrophobic
pores, this term reduces to an integral of the geodesic curvature on
the pore boundary~\cite{kamien}. A similar situation is encountered in
the case of proteins and other membrane inclusions, where the
orientation of the lipids at the membrane-inclusion boundary is
determined by the structure of the protein \cite{bhgdead1}. The change
in the Gaussian curvature in that case is 
\begin{equation}
\Delta {\cal H}_{\rm{gauss}} = 2\pi\bar{\kappa}\left(\cos\Omega-1\right),
\end{equation}
where $\Omega$ is the contact angle along the boundary.

The case of a hydrophilic pore, on the other hand, is more subtle.
Each of the two monolayers making up a bilayer can have an associated
Helfrich energy, and the two monolayers are joined at the pore
boundary. This, however, does not present a topological boundary and,
therefore, there is no integral of the geodesic curvature. The opening
of a hydrophilic pore results in a contribution to the Hamiltonian due
to the change in topology: The bilayer is topologically a sphere (if
we assume both bilayers are linked at the outer radius $L_p$), which
changes genus upon the opening of a pore, becoming a torus. For a
manifold without boundary, the Gauss-Bonnet theorem ensures that the
total Gaussian curvature is a topological invariant and thus measures,
to some extent, the global properties of the membrane. The change in
Gaussian curvature energy due to the formation of a hydrophilic pore
is given by
\begin{equation}
\Delta {\cal H}_{\rm{gauss}} = -4 \pi \bar{\kappa}.
\label{gausscurvature}
\end{equation}

The Gaussian curvature modulus may take both positive or negative values,
hence leading to either an increase ($\bar{\kappa}<0$) or decrease
($\bar{\kappa}>0$) in the energy upon the opening of a pore. Strict
comparison with classical nucleation theory (Eq.{\ref{nucleation}) is
possible only for $\bar{\kappa}=0$ and, therefore, we will restrict the
following discussion to this special case. The Gaussian curvature term can
be interpreted as an additional contribution that lowers or raises the
free energy depending on the sign of $\bar{\kappa}$. This contribution is
independent of the pore size. It will influence the probability of opening
a small nucleation pore, but will have no effect on the size to which
(meta)stable long-lived pores grow.

\subsection{Membrane fluctuation energy}

The first two terms of the Helfrich Hamiltonian (\ref{eq:HelfrichEnergy})
are local in character. Therefore, the above argument regarding the
absence of boundaries in a porous membrane fails, and the pore can be
treated as if representing the inner membrane boundary. In order to study
the statistical mechanical behavior of the membrane, we define a
coordinate system $(r,\theta)=(r_0\leq r\leq L_p,0\leq\theta<2\pi)$ in which
the pore is described by a curve of constant $r=r_0$:
\begin{eqnarray}
\label{eq:Embedding}
\textbf{X}(r,\theta) &=& [r \cos(\theta)+\eta_x(r,\theta)] \hat{x} +
  [r \sin(\theta)+\eta_y(r,\theta)] \hat{y}\nonumber\\
& & + h(r,\theta) \hat{z}.
\end{eqnarray}
The function $h(r,\theta)$ represents the height of the membrane above
some flat reference plane. The function $r \cos(\theta) \hat{x} + r
\sin(\theta) \hat{y} + \vec{\eta} (r,\theta)$ is a mapping from
coordinates $(r,\theta)$ in which the membrane pore will be circular,
to points in three-dimensional space in which the pores will have an
arbitrary shape (see fig.~\ref{fig:embedding}). Thus,
$\vec{\eta}(r_0,\theta)$ is a measure of the deviation of the pore
from having a circular projected area, which we will assume to be
small. Our choice or $r_0$, which we will {\em define}\/ as the radius
of the quasi-circular pore, is made by equating the projected area of
the pore to $\pi r_0^2$. More specifically, we will require that
$\vec{\eta}$ satisfies the following boundary conditions (BCs):
\begin{eqnarray}
\label{etabcr0}
\hat{\theta}\cdot\vec{\eta}\left(r_0,\theta\right)&=&0\\
\int_0^{2\pi}d\theta\, \left[r_0+\hat{r}\cdot\vec{\eta}\left(r_0,\theta\right)
\right]^2&=&\pi r_0^2.
\label{eq:etabc2}
\end{eqnarray}
On the outer (frame) boundary we set
\begin{equation}
\vec{\eta}(L_p,\theta)=0.
\label{etabclp}
\end{equation}

\begin{figure}[h,t]
\begin{center}
\scalebox{.5}{\centering \includegraphics{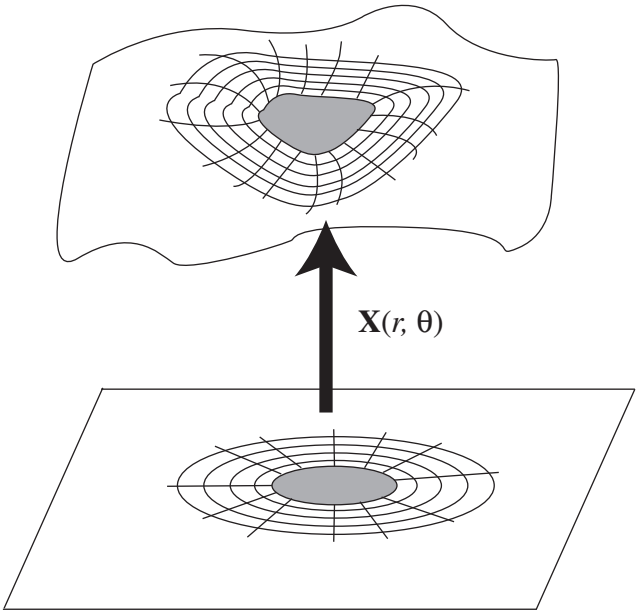}}
\end{center}
\caption{\label{fig:embedding} The mapping $\textbf{X} (r,\theta)$ 
  takes a flat reference plane containing a circular pore into a
  curved nearly flat membrane with a hole of nearly circular projected
  area.}
\end{figure}

With the embedding defined by Eq.(\ref{eq:Embedding}) we have, keeping
terms up to quadratic order in $\vec{\eta}$ and $h$, 
\begin{eqnarray}
g_{r r} &=& 1 + 2 \hat{r} \cdot \partial_r \vec{\eta} 
+ (\partial_r \vec{\eta})^2 + (\partial_r h)^2 \nonumber \\
g_{r \theta} &=& \hat{r} \cdot \partial_\theta \vec{\eta} 
+ r \hat{\theta} \cdot \partial_r \vec{\eta} 
+ \partial_r h \partial_\theta h + \partial_r \vec{\eta} 
\cdot \partial_\theta \vec{\eta}  \\
g_{\theta \theta} &=& r^2 + 2 r \hat{\theta} 
\cdot \partial_\theta \vec{\eta} 
+ (\partial_\theta \vec{\eta})^2 + (\partial_\theta h)^2
\nonumber
\end{eqnarray}
for the metric, thus giving us after a lengthy but straightforward
calculation
\begin{equation}
\label{eq:rootg}
\sqrt{g} \approx r \left[1+ \frac{1}{2}\left(\vec{\nabla}h\right)^2
+\vec{\nabla}\cdot\vec{\eta}+\vec{\nabla}\cdot\vec{\eta}_2\right]
\end{equation}
where 
\begin{equation}
\label{eta2}
\vec{\eta}_2=\left(\eta_x\partial_{\theta}\eta_y\right)
\left(\hat{r}/r\right)
-\left(\eta_x\partial_r\eta_y\right)\hat{\theta}.
\end{equation}  
One also finds that to lowest order the total curvature is given by
\begin{equation}
H =  \frac{1}{\sqrt{g}} \partial_\alpha \left( \sqrt{g} 
g^{\alpha \beta} 
\partial_\beta \textbf{X} \right) \approx \hat{z} \nabla^2 h
+ {\cal O}(h^2, \vec{\eta}^2).
\label{meancurvature}
\end{equation}

Substituting Eqs.(\ref{eq:rootg}) and (\ref{meancurvature}) into
Eq.(\ref{eq:HelfrichEnergy}) and keeping terms up to quadratic order
in $\vec{\eta}$ and $h$, we find
\begin{eqnarray}
{\cal H} &=& \sigma \pi (L_p^2-r_0^2)  \\
&+&\frac{1}{2}\int dr d\theta\, r \left[\sigma
\left(\nabla h\right)^2+ \kappa\left(\nabla^2 h \right)^2\right]
\nonumber \\
&+&\int dr d\theta\, r\sigma \left[\vec{\nabla}\cdot\vec{\eta}
+\vec{\nabla}\cdot\vec{\eta}_2\right]\nonumber.
\end{eqnarray}
The last integral in the above Hamiltonian can be converted into a
line integral by the application of the divergence theorem.  Using
Eqs.(\ref{etabcr0}), (\ref{etabclp}), and (\ref{eta2}) we arrive,
after some calculation, to the following form:
\begin{eqnarray}
\label{eq:QuadraticEnergy}
{\cal H} &=& \sigma \pi (L_p^2-r_0^2)  \\
&+&\frac{1}{2}\int dr d\theta\, r \left[\sigma
\left(\nabla h\right)^2+ \kappa\left(\nabla^2 h \right)^2\right]
\nonumber \\
&-&\frac{\sigma}{2} \int_0^{2\pi} d\theta\,\left[\hat{r}\cdot\eta\left(
r_0,\theta\right)\right]^2 - \sigma r_0 \int_{0}^{2 \pi} d\theta\, 
\hat{r}\cdot\eta(r_0,\theta) \nonumber.
\end{eqnarray}
An important feature of Eq.(\ref{eq:QuadraticEnergy}) is that,
\textit{to quadratic order}\/, the dependence of the Hamiltonian on
$h(r,\theta)$ and $\vec{\eta}(r,\theta)$ decouples completely.  Also,
notice that Eq.(\ref{eq:QuadraticEnergy}) depends only on the boundary
value of $\vec{\eta}$. This is a consequence of the fact that away
from the boundary, $\vec{\eta}$ is merely a transformation of
coordinates under which the Helfrich Hamiltonian should be invariant.
We will henceforth use the scalar function $\eta(\theta)$ to denote
the the boundary values of the mapping $\vec{\eta}(r,\theta)$, i.e.,
$\eta(\theta)\equiv \hat{r}\cdot\vec{\eta}(r_0,\theta)$.  Using BC
(\ref{eq:etabc2}), we find out that the last two
terms of Eq.(\ref{eq:QuadraticEnergy}) cancel each other.  This leaves
us with ${\cal H}=\sigma \pi (L_p^2-r_0^2)+{\cal H}_h$, where
\begin{equation}
{\cal H}_h\equiv \frac{1}{2} \int dr d\theta\, r
\left[ \sigma \left(\nabla h \right)^2 + \kappa
  \left(\nabla^2 h \right)^2 \right].
\label{eq:surfaceterm}
\end{equation}

The Laplacian in the height-dependent Hamiltonian
(\ref{eq:surfaceterm}) requires that we have two BCs on each boundary.
On the outer boundary ($r=L_p$) we impose the BCs
\begin{equation}
h(L_p)=0,\ \ {\rm and},\ \ \nabla^2 h(L_p)=0.
\label{outerbc}
\end{equation}
The first BC corresponds to a membrane which is attached to a static
frame on its external perimeter. The second is obtained by considering
the discrete version of the Helfrich surface Hamiltonian and requiring
that in the continuum limit, the same equation describes the motion
of the boundary and bulk elements. The BCs on the inner boundary ($r=r_0$) 
are quite similar:
\begin{equation}
h(r_0,\theta)=H(\theta)\ \ {\rm and},\ \ \nabla^2 h(r_0)=0,
\label{innerbc}
\end{equation}
with the only difference that the height is set to $H(\theta)$ rather
than vanishes. The vector
\begin{equation}
\label{poreline}
\mathbf{Y}(\theta)\equiv\mathbf{X}(r_0,\theta)=
\left[r_0+\eta(\theta)\right]\hat{r}+H(\theta)\hat{z},
\end{equation}
depicts the locus of the pore boundary in the 3D embedding space. Note
that in the case of a membrane inclusion of radius $r_0$, the second
BC in Eq.(\ref{innerbc}) should be replaced by $-\hat{n}\cdot
\vec{\nabla} h(r_0,\theta)=\partial h(r_0,\theta)/\partial
r=H'(\theta)$ where the contact slope, $H'$, depends on the geometry
and the tilt angle of the inclusion \cite{bhgdead1}.

We proceed in analyzing the area term by writing the height
function as $h=h_0+f$, where $h_0$ is the extremum of Hamiltonian
(\ref{eq:surfaceterm}), i.e.,
\begin{equation}
-\sigma \nabla^2 h_0+ \kappa \nabla^4 h_0 = 0,
\label{eq:h0}
\end{equation}
subject to the BCs that 
\begin{eqnarray}
&&h_0(L_p)=0,\ \ \nabla^2 h_0(L_p)=0,\nonumber \\
&&h_0(r_0,\theta)=H(\theta),\ \ \nabla^2 h_0(r_0)=0.
\label{bch0}
\end{eqnarray}
Eqs.(\ref{innerbc}) and (\ref{bch0}) imply that the function $f$,
which depicts the fluctuations around the equilibrium profile $h_0$
satisfies 
\begin{eqnarray}
&&f(L_p)=0,\ \ \nabla^2 f(L_p)=0,\nonumber \\
&&f(r_0)=0,\ \ \nabla^2 f(r_0)=0.
\label{bcf}
\end{eqnarray}
  
Hamiltonian (\ref{eq:surfaceterm}) can be thus written as
\begin{eqnarray}
\label{eq:helfrich2}
{\cal H}_h\left(h_0+f\right)=\int dr d\theta\, r \left\{ \frac{1}{2}
\left[ \sigma \left(\nabla h_0 \right)^2 +
\kappa \left(\nabla^2 h_0 \right)^2 \right]+\right. \\
\left[\sigma \nabla h_0 \cdot \nabla f
+ \kappa \nabla^2 h_0 \nabla^2 f \right]
+  \left.\frac{1}{2}\left[ \sigma \left(\nabla f \right)^2
+ \kappa \left(\nabla^2 f \right)^2 \right] \right\}&.&\nonumber
\end{eqnarray}
For the cross term (second term in ${\cal H}_h$) we obtain, upon
integration by parts,
\begin{eqnarray}
&&\int dr d\theta\, r  \left[\sigma \nabla h_0 \cdot \nabla f
+ \kappa \nabla^2 h_0 \nabla^2 f \right]= \nonumber \\
&&\int dr d\theta\, r
\left[ - \sigma \nabla^2 h_0 + \kappa \nabla^4 h_0 \right]f
- \nonumber \\
&& \int_0^{2\pi}d\theta\, \kappa \nabla^2 h_0 \frac{\partial f}{\partial r}
+\nonumber \\
&&\int_0^{2\pi}d\theta\, \frac{\partial \left[\kappa  \nabla^2 h_0 -\sigma  h_0
\right]}{\partial r} f,
\label{crossterm}
\end{eqnarray}
where the last two integrals in the above equation are performed on
the boundaries of the system. The boundary terms in
Eq.(\ref{crossterm}) vanish due to the BCs (\ref{bch0}) and
(\ref{bcf}), while the bulk term vanishes due to Eq.(\ref{eq:h0}).

Without the cross term in Eq.(\ref{eq:helfrich2}), the
height-dependent Hamiltonian takes the simple form ${\cal H}_h(h_0 +
f) ={\cal H}_h(h_0) +{\cal H}_h(f)$, where the energies associated
with $h_0$ (the equilibrium term) and $f$ (fluctuation term)
completely decouple.  Integrating both terms by parts twice, we find
expressions similar to Eq.(\ref{crossterm}), where $h_0$ is replaced
by $f$ (in the fluctuation term) or vice versa (equilibrium term). In
the former, the boundary terms vanish due to Eq.(\ref{bcf}) and we are
left with
\begin{equation}
{\cal H}_h\left(f\right)=\int dr d\theta\, r \frac{1}{2}
\left[ \sigma \left(\nabla f \right)^2 +
\kappa \left(\nabla^2 f \right)^2 \right].
\label{flucterm}
\end{equation}
In the latter, the bulk term is eliminated by virtue of
Eq.(\ref{eq:h0}). Considering the BCs on $h_0$ (\ref{bch0}), one can
easily find that
\begin{equation}
{\cal H}_h(h_0)= \frac{1}{2}\int_0^{2\pi}\! d\theta\, r_0H(\theta)
\frac{\partial \left(\kappa \nabla^2 h_0-\sigma h_0\right)}
{\partial r}\Biggm|_{r_0}.
\label{equilterm}
\end{equation}
In a manner similar to the last two terms in Eq.(\ref{eq:QuadraticEnergy}), the
last result demonstrates that contributions to the Hamiltonian due
to the pore can only appear through boundary (line) integrals.

\subsection{Pore line tension}

An additional contribution to the free energy is due to the line
tension of the pore, which arises from the curvature and packing of
the lipid molecules at the pore boundary (fig.~\ref{fig:inclusion_pore}).
The pore's shape is depicted by the curve $\mathbf{Y}(\theta)$
(\ref{poreline}), and the line tension energy is given by
\begin{eqnarray}
& &\Gamma \int_0^{2\pi} d\theta \sqrt{\left(d \textbf{Y}(\theta)/d \theta\right)^2} 
\simeq
\Gamma\,2 \pi r_0 + \Gamma \int_0^{2\pi} d\theta\, \eta(\theta) \nonumber\\
& & + \frac{\Gamma}{2 r_0} \int_0^{2\pi} d\theta 
\left[ \left(d H(\theta)/d\theta\right)^2+
\left(d \eta(\theta)/d\theta\right)^2
 \right], 
\label{lineterm}
\end{eqnarray}
where $\Gamma$ is the line tension coefficient, i.e., the edge
energy per unit length.
\subsection{The full Hamiltonian}

Collecting expressions (\ref{eq:QuadraticEnergy}), (\ref{flucterm}),
(\ref{equilterm}), and (\ref{lineterm}), we find that the Helfrich
Hamiltonian (excluding the Gaussian curvature term) can be written as
the sum of two terms:
\begin{equation}
{\cal H}={\cal H}_s+{\cal H}_l.
\end{equation}
The first term is the surface Hamiltonian associated with the membrane
fluctuations:
\begin{equation}
{\cal H}_s=\sigma \pi(L_p^2-r_0^2)+\frac{1}{2}\int dr d\theta\, r 
\left[ \sigma \left(\nabla f \right)^2 +
\kappa \left(\nabla^2 f \right)^2 \right].
\label{surfacetension}
\end{equation}
The second is the inner boundary term, consisting of the various
contributions to the energy due to the pore
\begin{eqnarray}
\label{linetension}
{\cal H}_l&=&\Gamma\,2 \pi r_0\nonumber\\
&+& \frac{1}{2}\int_0^{2\pi} 
d\theta\, \Biggl\{
\left[2\Gamma\eta+\frac{\Gamma}{r_0} 
\left(\frac{d \eta(\theta)}{d\theta}\right)^2\right]\\
&+&
\left[\frac{\Gamma}{r_0}\left(\frac{d H(\theta)}{d\theta}\right)^2+
r_0H(\theta)\frac{\partial \left(\kappa \nabla^2 h_0-\sigma h_0\right)}
{\partial r}\biggm|_{r_0}\right]\Biggr\}
\nonumber.
\end{eqnarray}

\section{Surface tension}
\label{section_surface}

In calculating the contribution of the surface tension term in
Hamiltonian (\ref{surfacetension}) to the free energy, we follow the
procedure described in our previous manuscript on membrane inclusions
\cite{bhgdead1}. We start by expanding the function $f$ in a series of
eigenfunctions $f_{m,n}(r)$ of the operator ${\cal L}\equiv-\sigma
\nabla^2 + \kappa \nabla^4$: $f(r,\phi) = \sum_{m,n} {\cal
  A}_{m,n}f_{m,n}(r) e^{i m \phi}$. The functions $f_{m,n}(r)$ can be
written as the linear combination of the Bessel functions, $J_m(r)$
and $Y_m(r)$, of the first and second kinds of order $m$, and the
modified Bessel functions of the first and second kinds of order $m$,
$K_m(r)$ and $I_m(r)$:
\begin{eqnarray}
f_{m,n}(r)&=&A J_m(\lambda_1^{m,n} r) + B Y_m(\lambda_1^{m,n} r)\nonumber \\
&+& C K_m(\lambda_2^{m,n} r)  + D I_m(\lambda_2^{m,n} r), \nonumber
\end{eqnarray}
where the $\lambda_i$ ($i=1,2$) are the positive solutions of
$(-1)^{i+1} \sigma (\lambda_i^{m,n})^2 + \kappa (\lambda_i^{m,n})^4 =
\mu_{m,n}$, and $\mu_{m,n}$ is the eigenvalue corresponding to the
function $f_{m,n}(r)$: ${\cal L}f_{m,n}(r)=\mu_{m,n}f_{m,n}(r)$.

Applying the BCs (\ref{bcf}) at $r_0$ and $L_p$, we derive the
eigenvalue equation
\begin{equation}
J_m(\lambda_1^{m,n} r_0) Y_m(\lambda_1^{m,n} L_p)
- J_m(\lambda_1^{m,n} L_p) Y_m(\lambda_1^{m,n} r_0)=0.
\label{eq:eigenvalues}
\end{equation}
Although this eigenvalue equation is different from the one we had in
Ref.\cite{bhgdead1}, the asymptotic behavior of the eigenvalues is the
same. In the long wavelength limit, $\lambda_1^{m,n}r_0\ll |m|$,
Eq.(\ref{eq:eigenvalues}) reduces to the eigenvalue equation in the
absence of pores:
\begin{equation}
J_m(\lambda_{1}^{m,n} L_p)=0.
\label{nopore}
\end{equation}
This is a manifestation of the fact that modes with characteristic
lengths much larger than the pore radius are hardly perturbed by its
presence. In the opposite limit, $\lambda_1^{m,n} r_0 \gg |m|$, we
find that the difference between two consecutive eigenvalues saturates
to $\lambda_1^{m,n+1}-\lambda_1^{m,n}=\pi/(L_p-r_0)$, which is a
factor $L_p/(L_p-r_0)$ larger than in the case with no pore. The
physical interpretation of this result is that the pore acts like a
hard wall for modes with characteristic lengths much smaller than its
radius, reducing the effective linear size of the membrane for these
modes to $L_p-r_0$.

The insertion free energy associated with the surface Hamiltonian,
defined as $\Delta F_s(r_0)=F_s(r_0)-F_s(0)$, can be expressed by the
following sum \cite{bhgdead1}
\begin{eqnarray}
\label{eq:freeenergydiff}
&&\Delta F_s(r_0) \approx - \pi \sigma_0 r_0^2 \\
&& +\frac{k_BT}{2} \sum_{m,n} \ln \left[\frac{\sigma (\lambda_1^{m,n})^2
+ \kappa (\lambda_1^{m,n})^4}{\sigma (\lambda_{1,(0)}^{m,n})^2
+ \kappa (\lambda_{1,(0)}^{m,n})^4}\frac{L_p^2-r_0^2}{L_p^2}
\right],\nonumber
\end{eqnarray}
where the sum runs over the modes $n=0,1,\ldots,\sqrt{N_0}$, and,
$m=-\sqrt{N_0},\ldots,\sqrt{N_0}$ so that the total number of modes
$2N_0$ is proportional to the number of molecules forming the membrane
$N$, while $\lambda_{1,(0)}^{m,n}$ are the corresponding solutions of
the eigenvalue equation in the absence of pores (\ref{nopore}).
As in~\cite{bhgdead1}, Eq.(\ref{eq:freeenergydiff}) was derived by expanding the 
surface tension to quadratic order in $r_0/L_p$, which is valid for small pores 
$r_0\ll L_p$.
Analytical approximation of this expression is obtained by assuming
(based on our discussion of the asymptotic behavior of the eigenvalues
$\lambda_1^{m,n}$) that eigenvalues such that $\lambda_1^{m,n} r_0 <
\alpha |m|$ (long wavelength) are not affected by the pore, whereas
modes with $\lambda_1^{m,n} r_0 > \alpha |m|$ (short wavelength) grow
by a factor $L_p/(L_p-r_0)$. The dimensionless constant $\alpha$ is 
of the order of unity and its value will be determined later by exact 
numerical calculation of $\Delta F_s$. Using this ``step-function''
approximation for the eigenvalues $\lambda_1^{m,n}$, and evaluating
the sum in Eq.(\ref{eq:freeenergydiff}) as an integral, we obtain the
simple result (correct up to quadratic order in $r_0$) that
\begin{eqnarray}
&&\Delta F_s=F_s(r_0)-F_s(0)=
\nonumber \\
&&-\pi r_0^2\sigma+\frac{k_B Tr_0^2}{\alpha l_0^2}
\left\{2-\alpha-\left( \frac{l_0}{\pi\xi}\right)^2
\ln \left[ \left(\frac{\pi\xi}{l_0}\right)^2 + 1 \right] \right\}
\nonumber \\
&&\equiv -\pi r_0^2(\sigma+\Delta \sigma)\equiv -\pi 
r_0^2\sigma_{\rm eff},
\label{rentension0}
\end{eqnarray}
where $\xi\equiv\sqrt{\kappa/\sigma}$, and $l_0=L_p/\sqrt{N_0}$ is
a microscopic length cutoff which is of the order of the bilayer
thickness.  From the above equation, we identify the thermal
correction to the surface tension as
\begin{equation}
\Delta\sigma= \frac{k_B T}{\pi\alpha l_0^2}
\left\{ \alpha-2+\left( \frac{l_0}{\pi\xi}\right)^2
\ln \left[ \left(\frac{\pi\xi}{l_0}\right)^2 + 1 \right] \right\}.
\end{equation}

\begin{figure}[ht]
\begin{center}
\scalebox{.4}{\centering \includegraphics{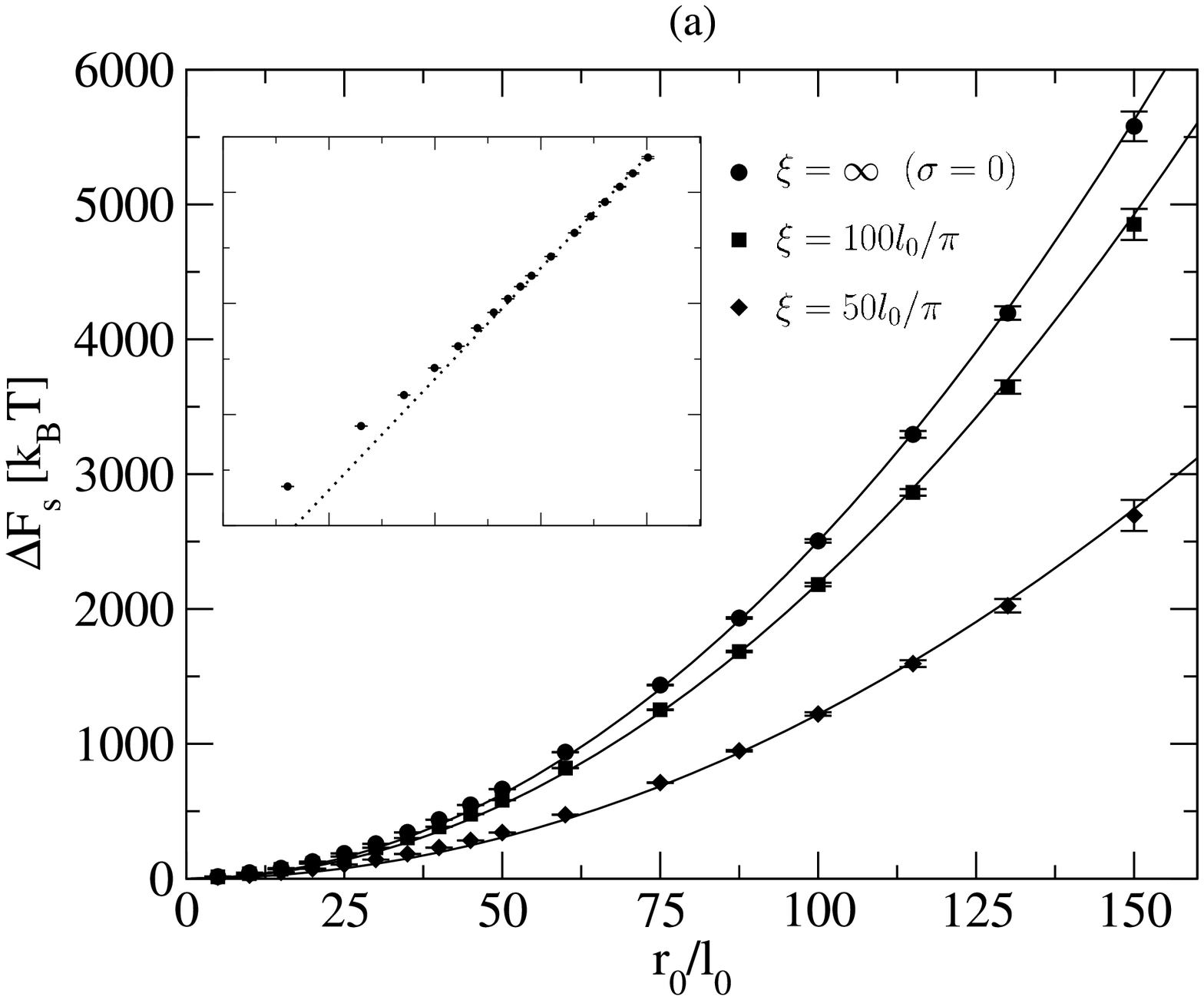}}
\vspace{0.3cm}
\scalebox{.4}{\centering \includegraphics{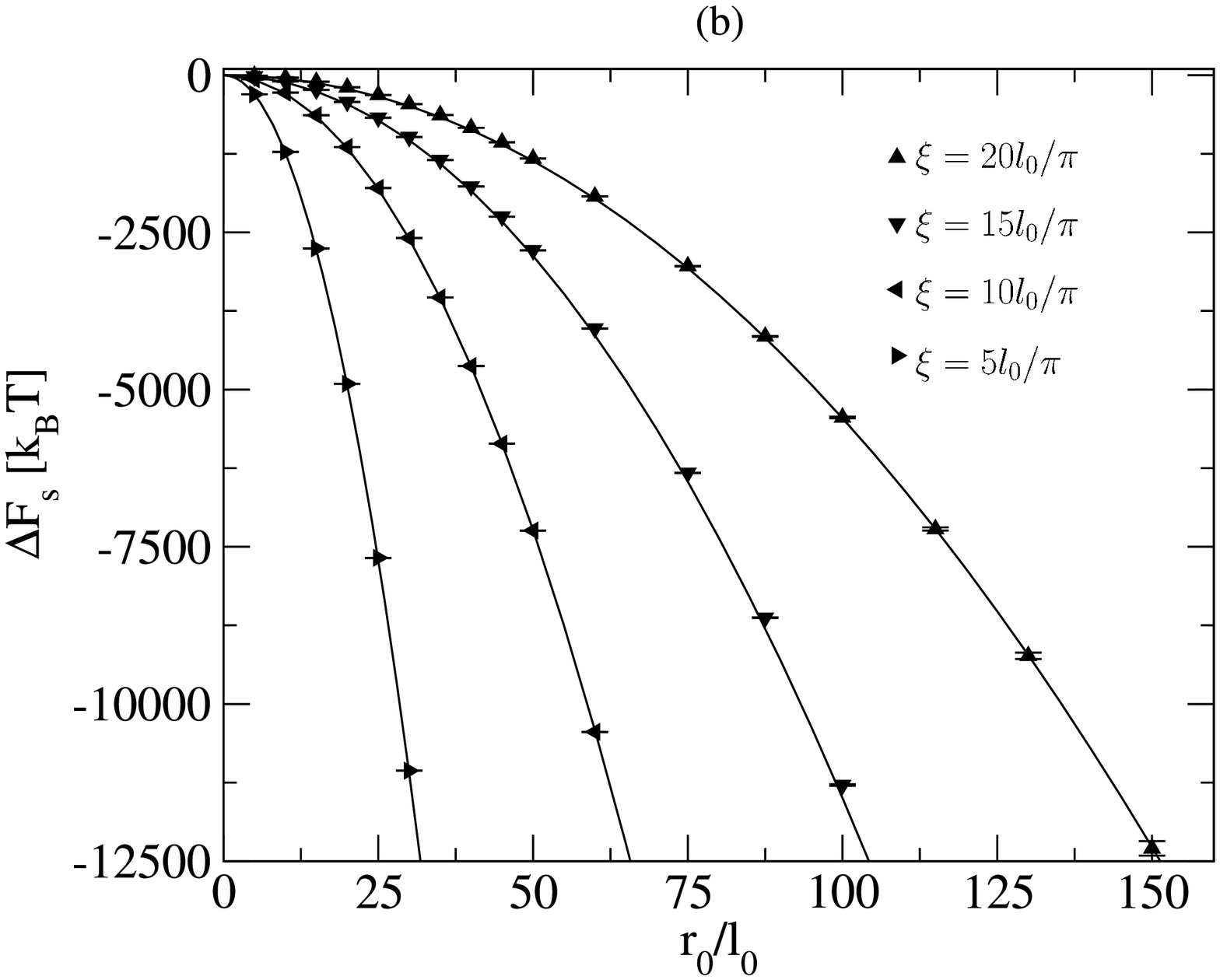}}
\end{center}
\vspace{-0.5cm}
\caption{The surface free energy $\Delta F_s$ as a function of the
inclusion's radius for $\kappa=10k_BT$ and various values of $\sigma$.
The inset to graph (a): a log-log plot of the numerical results for
$\sigma=0$. The slope of the straight dotted line is 2.}
\label{fig:fluct}
\end{figure}

In order to test the accuracy of expression (\ref{rentension0}), we
have numerically solved the eigenvalue equation (\ref{eq:eigenvalues})
and used the solutions to evaluate the sum in
Eq.(\ref{eq:freeenergydiff}). Numerical values of $\Delta F_s(r_0)$
(for $\kappa=10k_BT$ and various values of $\sigma$) are shown in
fig.~\ref{fig:fluct} (a)-(b). They have been extracted by
extrapolating the numerical results obtained for several values of
$750\leq N_0\leq 2000$ to the thermodynamic limit
$N_0\rightarrow\infty$. In the inset to fig.~\ref{fig:fluct} (a), the
results for $\sigma=0$ are replotted on a logarithmic scale, showing
that our prediction of a quadratic relation between $\Delta F_s$ and
$r_0$ is attained only for large (macroscopic) pores with $r_0\gtrsim
100l_0$ (the slope of the straight dotted line is 2). The discrepancy
between the numerical values of $\Delta F_s$ and
Eq.(\ref{rentension0}) in the small $r_0$ regime is due to the
significant contribution to the free energy of the crossover modes
$\lambda_1^{m,n} r_0\sim 1$ which is poorly calculated by the ``step
function'' approximation. The solid curves in fig.~\ref{fig:fluct}
(a)-(b) depict our analytical expression (\ref{rentension0}) for
$\Delta F_s$, with $\alpha$ determined by fitting the results for
large $r_0$ to Eq.(\ref{rentension0}). The value of $\alpha$ shows a
slight dependence on the surface tension varying from 1.60 for
$\sigma=0$ to 1.75 for $\xi=\sqrt{\kappa/\sigma}=5l_0/\pi$.

Our numerical and analytical results suggest that $\Delta \sigma<0$,
making the effective surface tension smaller than the bare surface
tension. Of particular interest is the fact, demonstrated in
fig.~\ref{fig:fluct} (a), that for weakly stretched membranes (large
$\xi$) the effective tension may be negative. In such a case the
effective surface tension would act to prevent, rather than
facilitate, the opening of a pore. For strongly stretched membranes
(small $\xi$) the dominant contribution to $\sigma_{\rm eff}$ is of
the bare surface tension. In this regime, the surface tension part of
the free energy is well approximated by the second term of
Eq.(\ref{nucleation}), i.e., $\delta F_s\simeq -\sigma\pi r_0^2$.

\section{Line tension}
\label{section_line}

In order to calculate the contribution of the line tension term in
Hamiltonian (\ref{linetension}) to the free energy, one needs to trace
out the fields $\eta$ and $h_0$ which are decoupled from each other.
Introducing the Fourier transform of the field $\eta$,
\begin{equation}
\eta(\theta)=\frac{d_{\rm dB}}{\sqrt{2N_1}}
\sum_{m=-N_1}^{N_1}
\tilde{\eta}_me^{im\theta}, 
\label{fouriereta0}
\end{equation}
where $d_{\rm dB}$ is the de-Broglie thermal wavelength. Making the
particular choice of $\tilde{\eta}_0$ that satisfies BC
(\ref{eq:etabc2}),
\begin{equation}
\tilde{\eta}_0 \simeq -
\frac{d_{\rm dB}}{\sqrt{2N_1}}\sum_{m \ne 0}\frac{|\tilde{\eta}_m|^2}{2 r_0},
\end{equation}
the corresponding Hamiltonian takes the form
\begin{equation}
{\cal H}_l^{\eta}=\frac{\pi d_{\rm dB}^2}{2N_1}
\sum_{\begin{array}{ll}\scriptstyle{m=-N_1}\\ \scriptstyle{m\neq 0} 
\end{array}}^{N_1}
|\tilde{\eta}_m|^2 \frac{\Gamma}{r_0} \left( m^2-1 \right).
\label{fouriereta}
\end{equation}
The $|m|=1$ modes are trivial translation modes which do not contribute
to the energy of the pore. The number of $m\geq 2$ modes is equal to
the number of microscopic degrees of freedom, namely the number of
molecules on the rim of the pore.  Since this number is proportional
to the perimeter of the pore, we can write
\begin{equation}
N_1\simeq b\left(\frac{r_0}{l_0}\right),
\end{equation}
where $b$ is a numerical factor of the order of unity.  Tracing over
the variables $\tilde{\eta}_m$ is straightforward, giving the free
energy
\begin{equation}
F_l^{\eta}=\frac{k_BT}{2}\sum_{|m| > 1}\ln\left[\frac{d_{\rm dB}^2 
\left(m^2-1 \right) \Gamma/r_0}{k_B T N_1}\right].
\end{equation}
If the number of modes is large $N_1\gg1$ (i.e., $l_0\ll r_0$) then
the sum in the above expression can evaluated as an integral, giving
\begin{equation}
\label{feta}
F_l^{\eta}\simeq 2\pi r_0\, \frac{bk_BT}{\pi l_0}\left[
\frac{1}{2}\ln\left(\frac{bd_{\rm dB}^2\Gamma}{k_BTl_0}\right)
-1\right].
\end{equation}

The contribution of the field $h_0$ to the line tension free energy is
also tractable. From the partial differential equation
(\ref{eq:h0}) and the BCs (\ref{bch0}), it is easy to show that (for
$\sigma>0$) $h_0$ can be written by the following mode representation
\begin{equation}
h_0(r,\theta)=\frac{d_{\rm dB}}{\sqrt{2N_1}}
\sum_{m=-N_1}^{N_1}
\tilde{h}_m\left(\frac{r_0}{r}\right)^{|m|}e^{im\theta}+
\tilde{h}_0\frac{\ln\left(r/r_0\right)}{\ln\left(r_0/L_p\right)}.
\label{fourierh0}
\end{equation}
Notice that $\nabla^2 h_0=0$ everywhere and not only at the
boundaries. Substituting expression (\ref{fourierh0}) in
Eq.(\ref{linetension}), one arrives to the following Hamiltonian
\begin{equation}
{\cal H}_l^h=\frac{\pi d_{\rm dB}^2}{2N_1}
\sum_{m=-N_1}^{N_1}
|\tilde{h}_m|^2\left(\frac{\Gamma}{r_0}m^2+\sigma|m|\right).
\label{fourierh}
\end{equation}
Assuming that $\sigma\ll\Gamma/r_0$ (the weak stretching regime), it
is easy to conclude that the resulting contribution to the free energy
$F_l^{h}\simeq F_l^{\eta}$, and thus
\begin{eqnarray}
F_l&=&2\pi r_0 \Gamma+F_l^{\eta}+F_l^{h}\nonumber\\
\label{feline}
&\simeq&2\pi r_0\left\{\Gamma+\frac{bk_BT}{\pi l_0}\left[
\ln\left(\frac{bd_{\rm dB}^2\Gamma}{k_BTl_0}\right)
-2\right]\right\}\\
&\equiv&2\pi r_0(\Gamma+\Delta \Gamma)\equiv 2\pi r_0\Gamma_{\rm eff}
\nonumber.
\end{eqnarray} 
We thus identify the thermal correction to the line tension of the pore
\begin{equation}
\Delta \Gamma=\frac{bk_BT}{\pi l_0}\left[
\ln\left(\frac{bd_{\rm dB}^2\Gamma}{k_BTl_0}\right)
-2\right].
\end{equation}
For phospholipid bilayers at room temperature $\Delta \Gamma$ is 
negative and is typically in the range of $ 10^{-7}-10^{-6}$ dyn.

\section{Discussion and Summary}
\label{section_summary}

The opening of a membrane pore has been traditionally regarded as an
energetically-driven process. According to this view, the surface and
line tensions are the forces driving, respectively, the opening and
closure of pores. The balance between these opposing forces creates a
nucleation barrier for the formation of long-lived pores, and requires
the opening of a sufficiently large hole at the initial stage. 

In previous studies, the role of thermal fluctuations has been limited
to facilitating the opening of a nucleation pore. The critical pore
size (\ref{criticalradius}) and the height of the barrier
(\ref{barrier}) have been determined from Eq.(\ref{nucleation}) for
the pore energy.  However, at nonzero temperature an entropic part
must be added to the Eq.(\ref{nucleation}).  To fill the gap in the
literature on the subject, we have calculated the thermal
contributions to the pore free energy associated with (a) the shape of
the boundary of the hole, and (b) the fluctuation spectrum of the
membrane.  Our study suggests that the pore free energy may be
expressed by an equation similar to (\ref{nucleation})
\begin{equation} 
F=\Gamma_{\rm eff}2\pi r_0-\sigma_{\rm eff}\pi r_0^2, 
\label{nucleation2} 
\end{equation} 
in which the bare surface and line tensions are replaced by effective
(renormalized) values. Typically, we find that $\Gamma_{\rm
  eff}<\Gamma$ and $\sigma_{\rm eff}<\sigma$, reflecting two opposite
tendencies. The decrease in the line tension reduces membrane
stability against pore formation. It reflects the larger configuration
space available to a membrane with a hole present. The decrease in the
surface tension, on the other hand, makes the formation of pores
harder in comparison to the zero-temperature case.  This effect
originates from the change in the spectrum of membrane fluctuation
occurring upon the opening of the pore and the resulting increase in
bending energy.

We can identify a number of different regimes of pore stability. For
tense membranes with positive effective line tension, we find the
standard regime of classical nucleation theory.
From Eq.(\ref{nucleation2}), we can identify the stability criteria
for the growth of large pores to be $r_0 >
\Gamma_{\rm{eff}}/\sigma_{\rm{eff}}$.  For membranes with low
surface tension, the effective surface tension will be negative. In
this regime, pores will increase the free energy for all radii and one
should not expect the formation of pores spontaneously as long as the
effective line tension is positive.  This regime is quite unlike
classical nucleation theory, where a nucleation barrier for pore
formation always exists.

In the theory of thermally activated poration, the nucleation rate of
critical pores depends strongly on the the free energy barrier $\Delta
F=\pi \Gamma_{\rm eff}/\sigma_{\rm eff}$, as $\exp(-\Delta F/k_BT)$.
The height of the nucleation barrier decreases as one approaches the
temperature at which the effective line tension, $\Gamma_{\rm eff}$
(\ref{feline}), vanishes.  Above this temperature the barrier
disappears and the formation of pores occurs spontaneously. The growth
of the pore will not be stopped as long as the surface tension remains
positive. If the rate at which the surface tension is relaxed is too
small, the membrane will rupture. However, we have demonstrated in
section \ref{section_surface} that membrane fluctuations renormalize
the surface tension, making the {\em effective}\/ tension negative
when the {\em applied}\/ tension is small. Therefore, we may have a
situation where both $\Gamma_{\rm eff}$ and $\sigma_{\rm eff}$ are
negative. In such a case, the free energy attains a minimum, not a
maximum, for $r_0=\Gamma_{\rm eff}/\sigma_{\rm eff}$, and a stable
pore of that radius will be spontaneously formed. The radius of such a
pore can be easily varied by several orders of magnitude by varying
the applied tension and thereby tuning the value of $\sigma_{\rm
  eff}$.

The pore size can be also varied by manipulating the magnitude of the
line tension, e.g., by the addition of colipids that modify the
ability of the lipids to pack at the edge region. Experiments in
bilayer lipids to which lysoPC and cholesterol were added
demonstrated that these molecules affect the line tension in opposite
ways, with the former decreasing the line tension
\cite{chernomordik2} and the latter increasing it
\cite{zhelev,karatekin}. Typical values of the line tension found
experimentally are in the range of our estimate of the thermal
correction $\Delta\Gamma$ or somewhat larger.  This demonstrates the
significance of thermal effects in determining the stability of
membranes against pore formation. Reexamining previous studies which
do not include temperature-dependent corrections is of particular
importance because the experimental determination of the line tension
$\Gamma_{\rm eff}$ is not direct. It is based on a suitable model
relating the line tension to other measurable quantities such as the
surface tension, pore radius, etc.
 
To better understand the effect of thermal fluctuations on pore
formation, it is necessary to go beyond the second order expansion of
the Helfrich Hamiltonian in $h$ and $\eta$. In the $\Gamma_{\rm
  eff}<0$ regime, the system gains free energy by having pores with
large perimeter, which means that holes with shapes that strongly
deviate from circular will be highly favorable \cite{shillcock_boal}.
By restricting our analysis to small values of the field $\eta$ (and
hence to quasi-circular pores only), we probably underestimate the
magnitude of the thermal correction to the line tension.  Moreover,
for negative values of the effective line tension the formation of
many pores is also likely to occur. The collective behavior of these
pores and the (membrane-mediated) interactions between them
\cite{bruinsma_pincus} are not well understood. More insight on the
pores' architecture and their dynamical evolution can be gained by
molecular-level studies and computer simulations.

\begin{acknowledgements}
We would like to thank P.~Pincus, P.~Sens, and K.~Katsov for
stimulating and useful discussions. This work was supported by the
NSF under Award No.~DMR-0203755.  The MRL at UCSB is supported by NSF
No.~DMR-0080034.
\end{acknowledgements}

\end{document}